\documentclass[rmp,aps,twocolumn,showpacs,superscriptaddress]{revtex4-2}
\pdfoutput=1
\usepackage[normalem]{ulem}
\usepackage{graphicx}
\usepackage{amsmath}
\usepackage{amssymb}
\usepackage{latexsym}
\usepackage{fancyhdr}
\usepackage{array}
\usepackage{amsfonts}
\usepackage{mathrsfs}
\usepackage{mathtools}
\usepackage{color}
\usepackage{verbatim}
\usepackage{physics}
\usepackage[caption=false]{subfig}
\usepackage{natbib}
\usepackage{enumitem}
\usepackage{bbold}
\usepackage[colorlinks=true,urlcolor=blue,citecolor=blue,linkcolor=blue]{hyperref}

\usepackage{float}
\usepackage[dvipsnames]{xcolor}
\usepackage{color,soul}

\begin{document}

\title{Against (unitary) interpretation (of quantum mechanics): removing the metaphysical load}
\author{Marek \.Zukowski}
\affiliation{International Centre for Theory of Quantum Technologies (ICTQT),
University of Gdansk, 80-308 Gdansk, Poland}

\author{Marcin Markiewicz}

\affiliation{International Centre for Theory of Quantum Technologies (ICTQT),
University of Gdansk, 80-308 Gdansk, Poland}

\begin{abstract} 
In June 1925 Heisenberg arrived at Helgoland/Heligoland island to escape a fit of hay fever. He returned with a sketch of a strange theory of the micro-world, which we now call quantum mechanics. This essay  attempts to present a look at this theory, which tries to return to the original insight of Heisenberg on what should be the essence of a theory of atomic realm: it must be a theory of the observable events, in which fundamentally unobservable quantities have no place. No ontological status is given to elements of the mathematical formulation of the theory. The theory is about our description of events in laboratories, probabilities of which are given by the Born rule.

Following  Bohr, these events involve macroscopic measuring apparatuses, and the accessible final events are classically describable. Information about the events is cloneable, as it is of a classical nature. 
 
The modern quantum theory of classicality is the decoherence theory.
It treats ``the pointer variable'' of measuring apparatus as an open system interacting with an environment consisting of all other ``zillions'' of degrees of freedom of the device, and anything coupled to it. Because such environment is uncontrollable we have no possibility to reverse measurements. 
The quantum mechanical measurement theory based on decoherence theory is reproducing the predictions of Born rule.

Notwithstanding, possibility of reversing measurements and  of application of Born rule in situations other than these which lead to  macroscopically observable events are features of   a modification of quantum mechanics which is called by its adherents {\em unitary quantum mechanics}.
 As its predictions, which go beyond quantum mechanics, are not testable -- we claim that unitary quantum mechanics in an  interpretation of quantum mechanics. As such it is metaphysics.\\
\end{abstract}

\maketitle

\section{Introduction}
Physics is a science. Thus a statement can be treated as its ``law" only if it agrees with our experience of the World/Nature (this includes our experiments, and observations which use scientific instruments). Statements which are fundamentally untestable are hypotheses which belong to metaphysics. Such are all interpretations of quantum mechanics, which attribute to its mathematical tools meanings that are beyond experimentally observable events, while not affecting quantum  predictions  of these events. 

{\em In this essay }we show  that  ``unitary quantum mechanics", which according to its followers leads to some interesting paradoxes, is an interpretation of quantum mechanics, based on hypotheses that are untestable. The (operational) quantum mechanics, which is the one tested in every quantum experiment, is free of these paradoxes. 

The root of ``unitary" vs. operational discrepancy is that the latter treats  the measurement process as irreversible, and in different answers to the question of what is described by the state vector.

 The clearest manifestation of this is the insistence of the supporters of ``unitary quantum mechanics" that measurements can be ``in principle undone". ``Unitarists" also try to avoid the postmeasurement state vector collapse at any cost, including no attempt to describe it, but still accept the Born rule as a calculational tool.
 
 Modern understanding of the collapse postulate is via the decoherence theory applied to quantum measurement. It allows to replace Copenhagenish intuitions about classical treatment of the laboratory apparatuses by analysis showing emergence of the apparatuses' classical-like behavior via decoherence due to the interaction with  zillions of degrees of freedom describing the atomic/quantum structure of the  devices and their environment. This in turn can be shown to lead to the impossibility of reversing the apparatus-system interaction, which happens during any laboratory measurement process. The hypothesis that unitary interaction between system, pointer variable, detectors and environment leading to a measurement can be ``in principle undone" is untestable, as it is impossible to build a quantum simulator showing the possibility of controlling such a process for so complex systems. This is not a practical impossibility, but an absolute impossibility, as the environment is by definition uncontrollable. {\em Ipso facto}, the hypothesis of ``in principle possibility of undoing measurements" belongs to metaphysics, as it is untestable. In the case of predictions of factual events in the laboratories the ``unitary" quantum mechanics agrees with the operational one. It shares this property with all interpretations of quantum mechanics which do not affect its predictions.  Metaphysics begins when one requests that quantum mechanics should be  more than a mathematically formulated theory which predicts future observable events of a certain class  basing on events observed earlier (of the same class).




\subsection{Quantum mechanics: origins and interpretations}

Quantum mechanics is a set of laws of physics which was postulated to describe experimental observations and tests concerning Microworld.  It was originally constructed to understand, among other phenomena,  the physics of black body radiation, to understand the photoelectric effect, and to understand stability of atoms and their spectra. By ``understanding" we mean finding a common set of principles/laws which give all these phenomena, and many more, as their predictions.

Most importantly quantum mechanics {\em was not derived starting from earlier theories}, it was guessed. Earlier intuitions behind the Bohr model were discarded. The Heisenberg matrix mechanics, exposed in the famous 1925-1926 trilogy \cite{Heisenberg25, Born25, Born26},  was an act of desperation, in the case of which the hay fever of Heisenberg played an important role. Dirac's canonical quantization is a rule of thumb, which has a heuristic justification, and is acceptable because it works. Schroedinger's wave mechanics was an attempt to find a dynamical equation for the waves related with particles, which constitute the matter, which were postulated by de Broglie on purely intuitive grounds of an analogy with photons. 

Everybody knows that. But we repeat the above because of the confusion concerning ``interpretations of quantum mechanics". Of course, we talk here about something that could be named ``over-interpretations", as obviously one must always relate mathematical objects of a  physical theory with observable events. This could be called an interpretation of the mathematical notions used in the theory. We assume here that this must be absolutely limited to relation with events in the laboratory, or observational facts. Occam Razor must be mercilessly used\footnote{The reader willing to confront with the entire landscape of different interpretations of quantum mechanics can find their categorization in \cite{Cabello.17}, under the significant title ``Interpretations of quantum theory: A map of madness''.}.

The mathematical and conceptual form of quantum mechanics, so different from classical physics, leaves minds of many of us troubled. For some it is an unacceptable theory. However, one of us recollects a dictum by a philosopher of science (perhaps Ernst Mach, but we were not able to confirm this\footnote{We were able to find the following related statement by Einstein (when discussing Mach's legacy): {\em Concepts that have proven useful in ordering things can easily
attain an authority over us such that we forget their worldly origin and take them
as immutably given... It is not at all idle play when we are trained to analyze the
entrenched concepts, and point out the circumstances that promoted their justification and usefulness and how they evolved from the experience at hand}. This is taken from \cite{deWaal20}, p. 71.}):  {\em concepts start to cause troubles, when one forgets their origin}.

Quantum mechanics, as every physical theory, is a mathematically formulated tool to predict future events (final stage measurement outcomes) basing on earlier events (initial stage preparation of quantum systems via earlier measurements, and an evolution stage). This must be remembered.

Concerning ``unitary" version of quantum mechanics mentioned in the title and the abstract, one can find its exposition e.g. in \cite{Deutsch.85}. 

\begin{itemize}
    \item 
Our principal aim is to show that the ``unitary quantum mechanics" is an interpretation, as it extends the range of quantum predictions   to only fundamentally unperformable gedanken experiments, no laboratory observable prediction is modified. 
\end{itemize}

\section{Predictive Quantum laws - mathematical formulation}

We shall present here the most concise formulation of the laws, with some comments of their relation to the modern formulation. That is, we shall concentrate on the ``pure state" case, and ``projective measurements". POVM's  are easily derivable from the above\footnote{The most important tool is in this case the Naimark dilation theorem.}, and so is  density matrix description of mixed states.
We shall discuss only the unitary evolution and the collapse postulate, as the modern concept of ``completely positive maps" is derivable from these.

The formulation which we shall initially present is standard, or  ``orthodox" with a ``shut up and calculate" bias. But we must remark, that it is not always stressed in standard texts, that it pertains to experimental situation which could be summed up by the stages of preparation, evolution and measurement. We shall limit our discussion to the ``first quantization''.

\subsection{Preparation, event-readiness and state tomography}
Initial stage of preparation,  or rather its result, is described by a ``state" vector $\ket{\psi}$ which is a normalized member of a Hilbert space $\cal H$ of a dimension which is equal to the maximal number of mutually perfectly distinguishable measurement results,  which are allowed for the class of quantum systems that we consider. If the system is electron's spin then $\textrm{dim } {\cal H} = 2$, etc. We either have a trusted device which was tested to emit systems which form  a finite  ensemble sharing the property of being equivalently prepared, or we make selective measurements, which do the job. In theory we relate with this procedure an infinite statistical ensemble of equivalently prepared quantum systems described by $\ket{\psi}$.

The postulates of quantum mechanics apply to a canonical gedanken situation of preparation -- evolution -- measurement\footnote{ The first papers were basically oriented toward understanding the quantum structure of the matter (harmonic oscillator, Hydrogen atom), Born introduced his rule when addressing a scattering process, which is definitely a preparation-evolution-detection experiment.}. As an extra to the usual presentation of this, we claim, and therefore assume, that in the canonical gedanken situation preparation must be \textit{event ready} (heralded), because only then it defines the ensemble {\em before} the final measurement stage.
Note that the event-ready character of a preparation procedure is meaningful only for a class of experiments in which the particles are neither created nor annihilated during the evolution stage, since only in such a case one can uniquely relate a detection event with a concrete preparation. Such experiments have a natural description in terms of \textit{first quantization}, without the need to introduce quantum field modes and field excitations. An emblematic example of such experiments is provided by the passive linear optics \cite{Pan12}, in which photon amplitudes are either beam-split or phase shifted, but no creation or annihilation happens at the evolution stage. Of course, one can describe such systems in terms of Fock space and field modes, however, this is not necessary and is just  a matter of convenience.

Thus, preparation in the canonical gedanken experiment is based on initial entanglement, and measurement of one system to define the state describing the other one (see e.g. \cite{ZZHE}).  Notice that many assume that preparation is not equivalent to measurement. This is usually stated like this. Take a situation in which  via a polarization filter only photons of a certain polarization pass; this is often claimed as the preparation procedure. But then, the experimental run is undefined until we detect such a photon in the measurement stage! In other words we create the ensemble by post-selection (enough ugly word to scare us...). Note that post-selection can introduce correlations, which are not in any causal relation with the preparation procedure, which may spoil correlation-based tests of non-classical nature of observed correlations, see e.g. \cite{Blasiak2021}. Quantum mechanics in its basic formulation is about preselected ensembles, although one can build within it the theory of post-selected ensembles.

Concerning a tested device that emits particles in a pure state: the testing is done via a procedure which in effect boils down to quantum state tomography of the output particles. Tomography can be understood as an application of the quantum measurement rules applying to the canonical experiment,  without any knowledge of the preparation of the ensemble. In the case of the theoretical statistical ensemble tomography can define/calculate the state describing such an ensemble. 

 Tomography works only for an ensemble, and is impossible to perform for a single run of an experiment because of the no-cloning theorem \cite{Wootters1982}. 
Thus, without metaphysics, the resulting reconstructed state based on a tomography is a property of the ensemble.

Sometimes, the tomographically reconstructed state describing a large ensemble of quantum systems emitted by a device can be close to pure, and in such a case we can sell it, or rather commercialize, as an emitter of systems described by a specific pure state.

If after the preparation stage all quantum systems of the ensemble described by $\ket{\psi}$ evolve in time in the same way, this is described by a unitary transformation $\hat{U}$, which transforms the state vector into $\ket{\psi_{final}}=U\ket{\psi}$. That is, it transforms the statistical ensembles, and most importantly this transformation is deterministic. Different preparations lead to different final statistical ensembles. The measure of their degree of distinguishability stays constant during the unitary evolution of the ensembles: 
$ |\braket{\psi_{final}}{\xi_{final}}|=|\braket{\psi}{\xi}| $.

\subsection{Measurement and Born rule}

The final stage measurement is usually called measurement of an ``observable". An observable is  represented by linear operator $\hat{O}=\sum_{l=1}^{d} r_l\ketbra{b_l}{b_l}$, where $r_l$ are (usually) real numbers which represent ``values" obtained in the measurement of the observable, and where the state vectors $\ket{b_l}$ form an orthonormal basis in $\cal H$. If a system survives the measurement process, and the result is $r_l$, and the laboratory measurement is arranged in such a way that its repetition on the system gives the same value $r_l$, then this constitutes a new preparation of an ensemble of systems that gave the result $r_l$, which is described by the  state vector $\ket{b_l}$. This is the "state collapse", which for some is an anathema.

{\em Is the collapse strange?} -- Note that it is common do describe the collapse in the following misleading way: 
 during the measurement process a collapse of the quantum state occurs randomly (with probabilities following Born rule), and this results in the observer obtaining a
random outcome. While in fact, as a result of an ideal measuring process, observer obtains a random outcome, and this forces the observer to change the wave  function (state) describing the {\em post-measurement} factual situation, to one which is consistent with the result. Why, because it is the result which defines the wave function (as a comprehensive description of the ensemble of systems that gave the specific measurement result).
 As Fuchs and Peres write: ``collapse is something that happens in our description of the system, not to the system itself" \cite{fuchs2000quantum}.
For an illustration see Fig. \ref{FIG1} and its caption. As pointed out by Streater \cite{Streater2007}, collapse is an effect which is not restricted to quantum theory, in fact it happens in classical probability as well (\cite{Streater2007}, p.59): ``{\it This is the collapse of the quantum state; it
is just Bayes’s formula for conditioning in the classical theory created by the
choice of complete commuting set. If quantum theory has any puzzle, this is
also a puzzle for classical probability.}''  On page 3 of the same book we can find: "{\em [...] it is evident that classical probability
is a special case of quantum probability, and Bayes’s rule is a special case of
the collapse of the wave packet."}

\begin{figure} \label{FIG1}
     \centering
         \includegraphics[width=0.99\columnwidth]{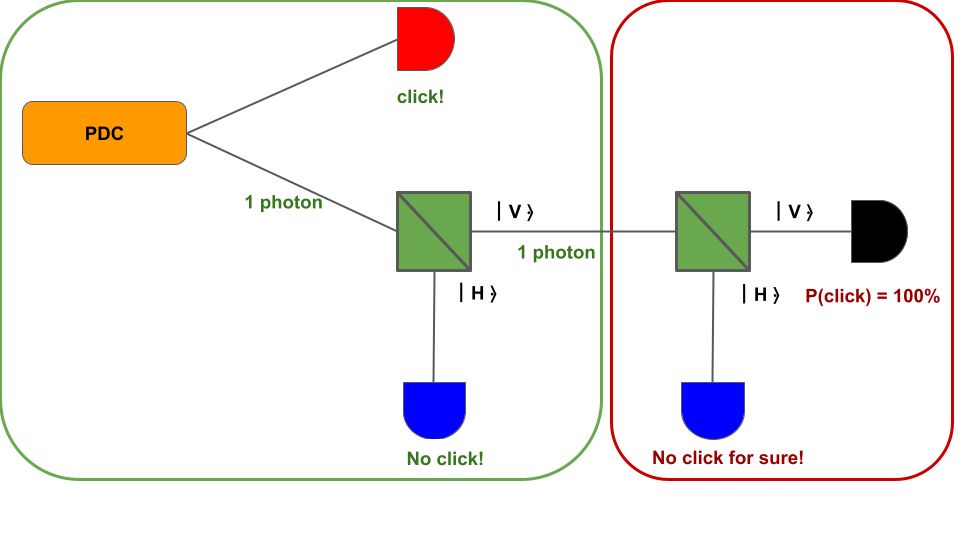}
        \caption{An illustration of a simple quantum mechanical operational situation, which involves two stages: preparation (green frame) and measurement (red frame). Preparation stage is based on the event-ready paradigm (is heralded). A Parametric Down Conversion (PDC) source, of type-II phase matched for producing polarization singlets (see e.g. \cite{Pan12}),  emits pairs of photons in coincidence, hence a click of the upper (red) detector indicates presence of a single photon in the lower path. The photon as a member of singlet pair has no defined polarization. This lower-path photon enters polarizing beam-splitter, which works as a polarization filter: if no-click is observed in the lower (blue) detector, the photon exits the polarizing beamsplitter via the exit for vertically polarized light. In this way one obtains a heralded/event-ready ensemble of identically prepared 
 vertically-polarized photons. The ensemble is represented in quantum mechanics by the ket $\ket{V}$. Measurement in a $\{\ket{H}, \ket{V}\}$ basis, realized behind the second polarizing beam-splitter, confirms the quality of preparation procedure: in the case of ideal detectors used both in the preparation stage as well as in the measurement stage, one has $100\%$ certainty that the photon would be detected by the upper (black) detector.\\ {\em Note that the green frame is a gedanken experiment illustrating that one must accept the collapse postulate as a part of quantum laws.} Assume that the PDC process is tuned to type-I, and thus the two exit beams of light are of the same polarization. If the polarization, described by say  $\ket{\xi}$, of the spontaneous down-conversion radiation heading to the polarization beamsplitter is different than $\ket{V}$, no-click event at the blue detector, together with a single photon click at the red (heralding) detector, imply that there is a photon in the exit channel described by  $\ket{V}$ heading to the final detection zone in the red box, and that there only the black detector is to click. In other words, the non-destructive measurement in the first box gave result $V$, and consequently we  ascribe to the exiting photon polarization membership in an ensemble described by the pure state $\ket{V}$.} \label{FIG1}
\end{figure}

At this stage quantum randomness shows up. Whereas evolution of the entire ensemble of preparations described by the initial state is deterministic, individual measurement outcomes are random events. Born's law, usually called Born rule, gives the probability of obtaining result $r_l$ as
$P(r_l|\psi_{final} )=|\braket{b_l}{\psi_{final}}|^2$. {\em Every quantum experiment for which we have a quantum mechanical prediction and  which ends with a measurement of an observable by an apparatus, or detection of particles in some counters is a test of this probability  rule.} Thus far, all such tests agreed with the rule, up to experimental errors, which are inevitable due to the finite number of repetitions of the experiment and imperfection of the devices.

And that's it. Nothing more.

\subsection{Remark on General Probabilistic Theories}
The described trio of a preparation, evolution and a measurement, which is from the very beginning inbuilt in the operational approach to quantum mechanics, inspired development of what is now referred to as Generalized Probabilistic Theories (GPT's), see e.g. recent review \cite{PLAVALA23}. In this very broad approach one tries to formulate a minimal mathematical framework for operational physical theory comprising the three stages. Then both classical as well as quantum mechanics (at least of finite-dimensional physical systems) can be embedded into the Generalized Probabilistic Theories' framework as two particular implementations of operational physical theories, see e.g. \cite{SELBY}. The Generalized Probabilistic Theories framework turns out to be very fruitful in describing limitations on information processing possible within assumed physical theory, which justifies its popularity in current foundational investigations. However it is worth to emphasize that it was quantum mechanics which motivated the development of the entire Generalized Probabilistic Theories framework.

\subsection{Where is entanglement?}
Entanglement which involves at least two systems, called by Schroedinger ``the essence of quantum mechanics", is often put  as a consequence of  one more quantum postulate. This is unnecessary, because if we have a Hilbert space of a non-prime dimension, e.g. $\textrm{dim } \mathcal{H} = d_{\cal A} d_{\cal B}$ where $d_{\cal A}$ and $d_{\cal B}$ are primes, one can easily show that it is homomorphic with a tensor product of a $d_{\cal A}$  and $d_{\cal B}$ dimensional Hilbert spaces. Etc.

\subsection{Where are specific interactions?}
Here we have addressed the ``kinematic" set of laws of quantum mechanics, without discussion of a specific dynamics, which is usually put in the form of the Schroedinger's equation, which gives the unitary transformation of the state: 
\begin{equation} \label{SCHR}
    i \hbar \frac{\partial}{ \partial t}\hat U(t, t_0)= \hat{H}(t)\hat U(t, t_0),
\end{equation}
with  $\hat{U}(t_0, t_0)= \hat I$. The interactions are described by a specific form of the Hamiltonian $\hat H(t)$. In the case of isolated systems the Hamiltonian does not depend on time.

\section{Copenhagen ``newspeak" in all that}
Note the following wording of the ``Orthodox" interpretation/presentation of quantum mechanics.

We have ``observables", not variables. This term was introduced to encode in the wording the fact that quantum measurement does not reveal values of certain variables which are defined for the given system  before the act of measurement/observation. Rather these values are {\em observed} at the instance of a measurement\footnote{Note, that Bell in his paper \cite{Bell90} is introducing a concept of ``beables", to express his skepticism toward quantum mechanics, and the orthodox parlance which we discuss in this section.}.  Upon very many  repetitions of the experiment on  equivalently prepared systems, the statistics of the observed values would approach Born rule.

 The original Schroedinger's dynamics, for a particle in space, is described as a dynamics of the wave \textit{function}, $\psi(\vec{x},t)$.\footnote{This was later extended to any state vectors $\ket{\psi(t)}$ describing any isolated system.} The word 'function' stresses that the thing is a mathematical tool, not something real. There is some confusion here, especially if one calls $\ket{\psi}$ the \textit{state of the system}, which for us is a misnomer. We do not talk about $\psi(\vec{x},t)$ as a wave field, a matter field, or whatever, despite the fact that from the point of view of pure mathematics this is a ``complex scalar field" 
 (a function with domain comprised of points in a geometrical space and values in the  field of complex numbers). A strong premise against treating $\ket{\psi}$ as a matter field is provided by delayed-choice-type experiments, initially proposed as thought experiments \cite{Wheeler78, WHEELER-2, Peres2000Del}, but which are nowadays testable experimentally, see e.g. \cite{Ma2012} for experimental test of delayed-choice entanglement swapping and  a recent review \cite{Ma2016}. The experiments of the delayed-choice kind  indicate that treating the wave function as a matter field forces one to accept some sort of retrocausal effects: \textit{``If one viewed the quantum state as a real physical object, one could get the paradoxical situation that future actions seem to have an influence on past [...]."} (from conclusions of \cite{Ma2012}).

\subsection{Remarks on what is $\ket{\psi}$}\label{PSI}
In the case of $\ket{\psi}$, as it was already argued above  we have a big trouble with nomenclature. 
The gut reaction is to call it ``the state of the quantum system" (in question). However, this immediately suggests that  $\ket{\psi}$ is a property of individual systems (like it is the case for a point mass in the canonical Hamiltonian formalism of classical mechanics, in the case of which $\vec{q}$ and $\vec{p}$ define the state, and are measurable properties of the point mass).

However, looking just at the formalism which expresses ``quantum laws" one sees that the role of $\ket{\psi}$ is that it is a descriptive tool which gives all {\em probabilities} of all possible measurements, via the Born rule 
$P(r_l|\psi)=|\braket{b_l}{\psi}|^2$. This formula holds for the measurement described by  the observable  $\hat O$, defined earlier. But this is a statement about the mathematics. One must remember that the formula was postulated by Born for probabilities of macroscopic events in the laboratory, related to counts in detectors, or spots on photographic plates, etc. That is, generally readouts of macroscopic measuring apparatuses. Within this realm it is both testable, and is being continuously tested in thousands of quantum labs.  Some try to apply it for events which are fundamentally untestable, but then paradoxes, i.e. contradictions emerge. The clinical example of that are the extended Wigner's Friend gedanken-experiments, \cite{Deutsch.85, Rovelli.96, Rovelli.21, Frauchiger.18, Brukner.18}. More about that further on.

Importantly $r_l$ plays here only a bookkeeping role. If $r_l$ is given a physical interpretation, e. g. it is a value of angular momentum, then it has specific values allowed for angular momentum. However it may be just the number tag of a detector in measurement station arbitrarily  ascribed by the experimenter.  This arbitrariness, allows even to ascribe complex values, see e.g. \cite{PhysRevA.55.2564}, or other non numeric symbols to measurement events. Therefore in quantum mechanics it is better to think in terms of projector observables, given in this case by  $\{\ket{b_l}\!\bra{b_l}\}_l$. 

Most importantly $\ket{\psi}$  is defined by the preparation process (essentially, a filtering measurement), plus perhaps the deterministic evolution described by a unitary transformation. It allows to calculate probabilities for all possible measurements and accounts for complementarity. This means that it does not give a joint distribution of outcomes of all possible measurements (in classical mechanics the joint probability of $\vec{q}$ and $\vec{p}$ always exists). Fully complementary observables are linked with mutually  unbiased orthonormal bases.

$\ket{\psi}$ represents an equivalence class of preparations of physical systems. However since it gives only probabilistic predictions and probabilities are defined for statistical ensembles, it is a description of a statistical ensemble of equivalently prepared systems. One may give other additional  attributes to $\ket{\psi}$, but this one is fundamental. Without it the quantum formalism makes no sense (see discussion in \cite{Bal70}).  The additional attributes are, e.g. the claim that $\ket{\psi}$ describes the individual quantum systems of the ensemble (Copenhagen interpretation in its most common form, \cite{Werner14}), that it represents directly unobservable relations in so-called Quantum Relational Space \cite{Horodecki25}, or that there are non-local hidden variables behind it, and therefore $\ket{\psi}$ is only an epistemic tool to describe observable effects {\em caused} by these (Bohmian interpretation). Interestingly, as pointed by R. Werner \cite{Werner14},  the  view that wavefunction describes a statistical ensemble of equivalently prepared systems was acceptable for Einstein himself: (cited from \cite{Werner14}, originally published in \cite{Schilpp59}): 
\begin{itemize}
    \item 
``\textit{One arrives at very
implausible theoretical conceptions, if one attempts to maintain the thesis that
the statistical quantum theory is in principle capable of producing a complete
description of an individual physical system. On the other hand, these difficulties
of theoretical interpretation disappear, if one views the quantum mechanical
description as the description of ensembles of systems. I reached
this conclusion as the result of quite different types of considerations. I am
convinced that everyone who will take the trouble to carry through such
reflections conscientiously will find himself finally driven to this interpretation
of quantum-theoretical description (the $\psi$-function is to be understood as the
description not of a single system but of an ensemble of systems)}".
\end{itemize}
As emphasized by Werner, Einstein's main motivation when constructing the EPR argument \cite{EPR.35} was the rejection of the ``individual system'' interpretation of $\psi$. Note, that most of the discussions on the EPR paper concentrate not on the interpretation of the ``$\psi$ function" but rather on the question of existence of (local) elements of reality, as additional hidden variables. As a matter of fact the question of existence of elemants of reality  was the topic of the celebrated Bell's paper \cite{Bell.64}.

We can summarize our position as follows (this is based on an excerpt from the original submitted version of our article \cite{Zukowski.21}):
\begin{itemize}
    \item 
    Interpretations
usually involve a specific understanding of the notion of
the quantum state. For us a quantum state is a theory specific description (in terms of, in general, density operators) of a statistical ensemble of equivalently prepared
systems, which allows for statistical (probabilistic) predictions of future measurements, via the Born rule. The
state describes an individual system only as a member of
such an ensemble. The theory itself is ``a set of rules
for calculating probabilities for macroscopic detection
events, upon taking into account any previous experimental information”, \cite{fuchs2000quantum}. Or if you like, one can use
E. P. Wigner’s statement “the wave function is only a
suitable language for describing the body of knowledge -
gained by observations - which is relevant for predicting
the future behaviour of the system” (a quotation from Wigner's article published in 1961, reprinted in \cite{Wigner.61}). Note that all
internally consistent interpretations (not modifications\footnote{Such modifications of quantum mechanics comprise, among other ones, e.g. nonlinear generalizations of quantum mechanics \cite{Zakharov1974, Bialynicki76, Czachor96}, and objective collapse models \cite{Ghirardi86, Ghirardi90, Diosi89, Penrose1996}.})
of quantum mechanics agree with the above. They only
add some other properties to the quantum state, or to
individual members of the ensemble (systems), without
any modification of the calculational rules of quantum
theory (based on the statistical ensemble approach).
\end{itemize}

In all experiments for which we have well defined quantum predictions we test the quantum formalism, but this is done via testing the Born rule. As all experiments have a finite duration, their raw results form finite ensembles of data. To estimate the probabilities all experimenters use the relative frequencies of occurrence of results, $r_l$. Thus whether one likes it or not, whether one interprets probabilities as propensities, in practice one uses the frequentist approach to probabilities to test the probabilistic predictions. One tests in the laboratory the optimal betting strategies of Qubism \cite{Caves02} in the same way\footnote{Provided they are applied to real experiments, see further down in appendices the discussion about ref. {\cite{Qbism2020}}.}.  

The pure state vector $\ket{\psi}$ is used to describe the situation in which the preparation has no stochastic element.
  This means that the preparation is maximal informationally. If the preparation is in a form of a filtering, a second act of the same filtering in the ideal case does not make a further selection, hence no stochasticity appears in such a procedure, see Fig. \ref{FIG1}. In fact, the existence of a (theoretical) measurement procedure which upon repetitions of the experiment boringly gives always the same result, a specific $r_l$, is an if-and-only-if property of a pure state. 
  \begin{itemize}
      \item 
Because of all that discussion above, we shall try to consequently call $\ket{\psi}$ as ``the state describing the system", instead of ``the state of the system", and sometimes even ``the ket describing the system", or other phrases which seem equivalent for us. 
  \end{itemize}

\section{Whose knowledge?}
\label{sec:teleport}
This is the question ... for some, e.g \cite{Bell90}. 
The knowledge about the preparation procedure  obviously belongs to the experimenter preparing the systems. It defines the ket $\ket{\psi}$. The experimenter can pass it to other actors or spectators, and even write it down to a manuscript, to inform Humanity. Thus the knowledge about the preparation $\ket{\psi}$ belongs to anyone who knows it. In our times information about the preparation procedure can be fed into some automatons, which would perform some operations (evolution, measurements). If we monitor the process the information can be always transformed into our knowledge.

\begin{figure}
     \centering
\includegraphics[width=0.99\columnwidth]{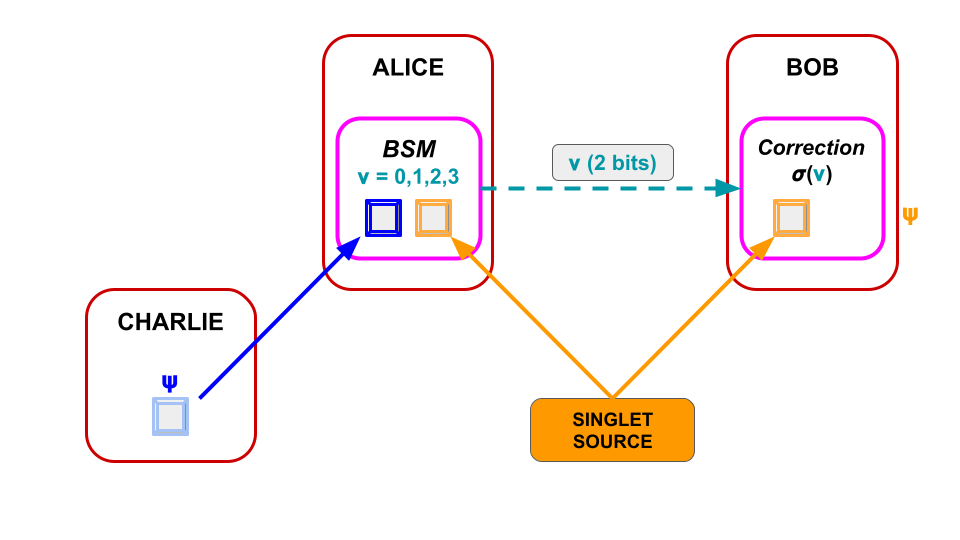}
        \caption{Schematic representation of the (qubit) teleportation protocol. A detailed description is in the main text. We just list here who knows what. Charlie knows that the particles which he sends ``carry" qubits, which are members of a statistical ensemble described by $\ket{\psi}\in \mathcal H^{(2)}$. Alice knows the construction of their teleporter and the protocol of the teleportation. After a run of the experiment she knows the result of her Bell State measurement, $\nu$. Bob in each run initially knows the construction of the teleport and the protocol, and nothing more. Only when he receives a two bit non-superluminal  message $\nu$ from Alice he can perform the protocol unitary transformation on his  particle from the pair carrying two-qubit singlet, to be sure after that his qubit is a member of an ensemble described by $\ket{\psi}$. But he does {\em not} know $\ket{\psi}$. Only a correlation was established, but local knowledge does not allow Bob to know $\ket{\psi}$. He must resort to tomography, and it is an operation on a statistical ensemble. However, Charlie knows that, if Alice and Bob followed precisely the protocol, and nothing malfunctioned, Bob's qubit belongs to an ensemble described by $\ket{\psi}$. }\label{fig:tele}
\end{figure}

\textit{The knowledge belongs only to those who know it.} Knowledge of different agents may differ.
Take the quantum teleportation process, see Fig. \ref{fig:tele}. As it is well known, only important stages will be discussed. We shall introduce here three characters: Charlie who sends a qubit  to Alice, who in turn preforms, on it  and on another qubit from a two-qubit  singlet state, a Bell-state measurement. The second qubit of the singlet can be manipulated in a distant laboratory by Bob. Charlie knows that the preparation of his qubit is describable by $\ket{\psi}$. Alice after her action knows that Bob's qubit is for her described as prepared in such a way that this act of preparation is described by $\sigma_{\nu}\ket{\psi}$, where $\nu=0,1,2,3$ are numbers which according to the protocol she ascribes to the Bell-state-measurement results\footnote{Symbol $\sigma_{\nu}$ represents Pauli operators, which are not only Hermitian, but also unitary.}. Alice and Bob do not know $\ket{\psi}$. Bob knows that his qubit is a sub-system of a pair of qubits which is a member  the singlet ensemble, and thus for him it is  \textit{initially} effectively locally described as belonging to an ensemble related with the  maximally mixed state (in other words he knows... nothing about his qubit). Only if Alice sends him the value of $\nu$ and he performs the unitary transformation $\sigma_{\nu}$, he knows that his qubit can be described as a member of equivalence class (statistical ensemble) of qubits which results from preparation by Charlie. But, he does not know that it is described by  $\ket{\psi}$! The information sent by Alice, does not allow him to get a description revealing some traits of $\ket{\psi}$, because the Bell state measurement (BSM) in each run of the experiment does not reveal any information about the ensemble prepared by Charlie (more strongly, for Alice, who has no access to Bob's qubit, it completely erases it).

Somebody who does not know the preparation procedure has to resort to the state tomography. This is done on a statistical ensemble (in theory), or (in the lab) on  a set  of statistically relevant number of experimental runs. In the case of a {\em perfect} single experimental run one is only able to find out which kets definitely did not describe the preparation (namely, all those which are orthogonal to $\ket{b_l}$ associated with the obtained result $r_l$). Note that, if Charlie prepares for each run of the teleportation experiment completely randomly chosen states, the teleportation would work in each run, but a possible post-teleportation tomography by Bob would be useless.

\section{Copenhagenish approach to understanding the measurement}

The essence of Bohr's approach to measurement is his insistence that it involves {\em classical, macroscopic apparatuses}. They produce classical information about the results. Note that such information is transferable by classical means, and thus in opposition to what we now know about its quantum counterpart \cite{Wootters1982} it is ``cloneable". This was postulated by Bohr in order to have clear link between our macroscopic observations with the quantum laws, esp. the Born rule. Most importantly, there is no experiment or observation which falsifies Bohr's approach. Still, to many it seemed to be sweeping a problem under a rug, or a kind of phenomenological approach.\footnote{And one could say that indeed it was so {\em before} the decoherence via interaction with uncontrollable environment theory of measurements.} 

The first important step toward quantum theory of measurement was von Neumann's analysis of what is now called {\em pre-measurement} \cite{vN.32}. It is the first stage of the measurement of  an  observable. We shall assume that it is $\hat{O}=\sum_lb_l \ketbra{b_l}{b_l}$.  Via a suitably tailored interaction, i.e. governed by a suitable Hamiltonian and of proper time duration, a certain different degree of freedom of the system or the device called  the pointer variable,  $P$, gets entangled with the eigenstates of the observable, and the description of the pair evolves in the following  way:

 \begin{equation} \label{VON-NEUMANN}
     \ket{\psi}_s\ket{neutral}_P \rightarrow
     \sum_lc_l\ket{b_l}_s\ket{r_l}_P, 
\end{equation}
where the subscripts denote the system and pointer, and $\ket{\psi}_s =
     \sum_lc_l\ket{b_l}_s$. Pointer starts in position ``neutral" and $r_l$ are "positions of the pointer" indicating the measured value. All that, we suggest, must and can be understood as an evolution of the description of a statistical  ensemble.

     A remark. Von Neumann used a bit different wording, but what we show above is the current understanding of his description: the process shown in (\ref{VON-NEUMANN}) is called pre-measurement. For him the evolution was $\ket{\psi}_s\ket{neutral}_A \rightarrow
\sum_lc_l\ket{b_l}_s\ket{r_l}_A,$ where kets with the subscript  $A$ are quantum states describing "apparatus". This was a bold step beyond the ruling Copenhagen orthodoxy, which at this time stressed the classical nature of measurement devices. As we shall argue further, the macroscopic nature of measuring devices, which leave readable traces in their internal and general environment, which are signaling the result of a measurement, allows one to return to the Copenhagen standpoint via the quantum theory of the classical, that is the decoherence theory, see e.g. \cite{Zurek1991}.
 
\begin{figure} 
     \centering
\includegraphics[width=0.99\columnwidth]{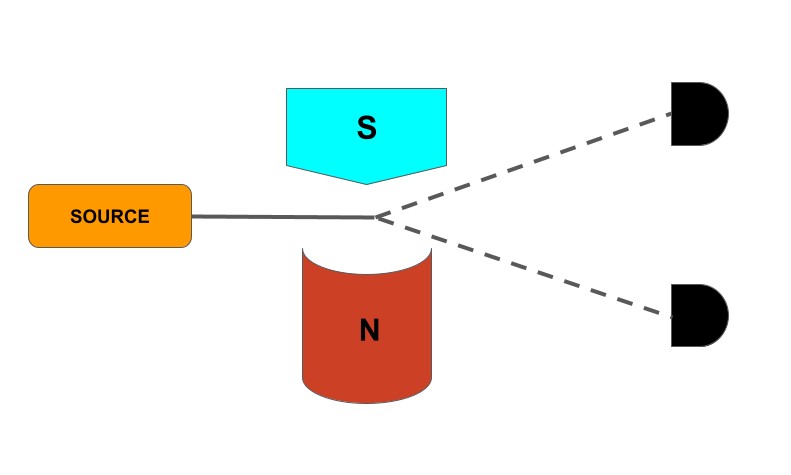}
        \caption{An emblematic example of a pre-measurement process takes place in a Stern-Gerlach device, which entangles particle's spin with its path. The latter one becomes the pointer variable. Detection of a particle by a detector  in  a specific path signals the associated spin value. }\label{fig:sg}
\end{figure}

 An iconic example of such interaction is the pre-measurement which takes place in a Stern-Gerlach device which entangles particle's spin with its path, for an illustration see Fig. \ref{fig:sg}. The latter one becomes the pointer variable. Detection of a particle by a detector  in  a specific path signals the associated spin value. Von Neumann  stresses irreversibility of the detection process, and that the other its feature is amplification of the initial quantum event to the level at which it can be  observed macroscopically. In quantum optics  the iconic amplification devices are avalanche photo-detectors: initial ionization produces a free electron  which is accelerated in the electric field inside the device, and in a collision ionizes another atom, and so on, till a macroscopically registrable electric current forms.

 Von Neumann overtly introduced the collapse postulate, which is unacceptable to some. However, a moment of thought leads us to the following. Quantum mechanics gives only probabilistic predictions for measurement results. Probabilistic theories describe statistical ensembles, but on the other hand ``trials". E.g. we have clear probability assignment for the results of throwing a dice, but in an individual trial only one of the six possible results occurs. Any probabilistic theory has ``trials" as its structural element, they form its empirical essence.
 Thus, the very nature of quantum (probabilistic) prediction assumes that in individual runs of an experiment only one specific event occurs.\footnote{Such an event might be collective, comprising of several ``sub-events" like e.g. detection of left-circularly polarized photon by Alice and horizontally polarized photon by Bob in a Bell experiment. In other words it is a collection of all events at the measuring apparatuses during a single run of the experiment.} The runs of an experiment are its trials. As such an event $r_l$, when used at the preparation stage based on filtering, defines an ensemble of equivalently prepared systems $\ket{r_l}$, we see here that this filtering selection is in fact the state collapse $\ket{\psi} \rightarrow \ket{r_l}$, understood as a change in the quantum  description due to the change of associated statistical ensembles: from the initial ensemble, described by $\ket{\psi}$, to the (sub)ensemble $ \ket{r_l}$ selected via acts of observations of the result ${r_l}$.

 \subsection{Quantum measurement theory based on decoherence}

 The von Neumann approach still did not answer the following two questions: 
 \begin{enumerate}
     \item 
 how the {\em irreversibility}  and amplification to the macroscopic level
 emerge? 
 \item
 and in consequence how initial coherence  of the system in the measurement  is irreversibly lost.
 \end{enumerate}
 That is one needs a quantum mechanical explanation of the (non-unitary) transition:  
 \begin{equation} \label{DECOH}
   \sum_lc_l\ket{b_l}_s\ket{r_l}_P  \rightarrow \sum_l|c_l|^2\ketbra{b_l}_s\ketbra{r_l}_P,
\end{equation}
and its irreversibility. Decoherence theory gives the answer. The dynamics is unitary, but it involves also the environment, internal and external of the measuring devices, which gets entangled with system and a collective observable describing pointer. The final formula in (\ref{DECOH}) gives what we expect in the experiment: a strict classical correlation between eigen-kets related with specific values of the observable $\hat{O}=\sum_lb_l \ketbra{b_l}{b_l}$, and eigen-kets of pointer read-out positions. The probabilities of readouts are given by $P(r_l)=|c_l|^2$.

\subsubsection{The initial coherence: lost or not lost}
 In the case of the second question, to elucidate its importance for our discussion let us introduce a ``unitarist" argumentation. The evolution which entangles system with the pointer variable is unitary. Therefore, it is described by a certain unitary operator $\hat U_{int}$, of the following property: $\hat U_{int}\ket{b_l}_s\ket{neutral}_P 
=\sum_l c_l\ket{b_l}_s\ket{r_l}_P.$ However this is not a unitary transformation operator which describes e.g. the intellectual operation of a change of orthogonal basis, but instead an operator which is to describe the {\em evolution} of system and pointer leading  to the entanglement postulated in the  von Neumann's model (\ref{VON-NEUMANN}). As such the evolution is a solution of the Schroedinger equation (\ref{SCHR}).

A ``unitarist" argumentation is that we can apply $\hat U_{int}^{-1}$ and the coherence returns, as the system is back described by $\ket{\psi}_s$.

The quantum measurement theory based on decoherence allows us to understand how the irreversibility emerges, and how emerges the amplification of the measurement signal to the macroscopic level (the first question). This level is characterized by a classical behaviour of the coarse grained description with collective variables.  The basic aim of this paper is to give a new simple argument for the irreversibility. Not on the FAPP level (For All Practical Purposes, as described by Bell, in a rather derisory way,  in his famous article entitled ``Against 'measurement' '' \cite{Bell90}), but an absolute impossibility of constructing a device that is able to inverse the unitary interaction leading to the measurement process. Our presentation of the decoherence theory would be sketchy since our aim is to emphasize just the most important concepts. Reader interested in tracing the history of the theory of decoherence should refer to old papers by Zurek \cite{Zurek81, Zurek.82, Zurek.03}, whereas the  presentation from a nowadays perspective can be found in the recent publications \cite{ZUREK2022, ZUREK-book-2025}. It is worth mentioning that similar ideas, although \textit{not} under the name of ``decoherence'', can already be found in the 1951 textbook of Bohm \cite{Bohm51}, chapter 22, published just before his turn into what one now calls \textit{Bohmian mechanics}. One can find there a subsection ``Irreversibility of Process of Measurement and Its Fundamental Role in Quantum Theory''.

\subsubsection{All things are made of atoms}
The above famous statement by Feynman forces one to consider classical measurement apparatuses  as also made of atoms. Apparatuses contain around Avogadro number of atoms ($6 022 000 000 000 000 000 000 00$) or something near to it. Rather more than one millionth of the number, which is around $10^{17}$. However each atom has plenty of ``degrees of freedom", and is described by an infinitely dimensional Hilbert space. It interacts with other atoms, and importantly also with the electromagnetic field. To describe the entire pointer we must resort to (quantum) statistical description, as it is impossible to precisely control its microscopic state. In the case of Stern Gerlach experiment, Fig. \ref{fig:sg}, the pointer variable belongs to the measured particle. Firing of one of the detectors might be treated as a collective yes-no variable ``fired"/``not-fired", and is obviously irreversible. Beyond this our detailed description is both impossible and impractical. This is just like in the transition from  classical statistical mechanics to thermodynamics, in which the microscopic properties of individual atoms, or rather point masses, are replaced by a small number of macroscopically relevant parameters.   The rest of the device is effectively an internal environment of the device, and additionally this environment couples to an external environment, which includes the electromagnetic fields, the wiring, computer, display, and if one discusses Wigner's Friend also at least her senses, brain or part of it, but perhaps more. The motion of the  macroscopically accessible pointer variables, like its position and momentum, becomes due to the decoherence effectively classical, see e.g. \cite{HAAKE}.

\subsubsection{Sketches of quantum measurement theory based on decoherence}

First, a technical remark. Because of the enormous lack of our knowledge about the microscopic state describing the broad environment one could insist that it must be described by a mixed state $\varrho_E= \sum_{\lambda} p(\lambda )\ketbra{\xi_{\lambda}}{\xi_{\lambda}}_E$.
But the linearity of quantum dynamics of density matrices allows us to resort to considerations which 
start with a pure state describing the environment, to be denoted $\ket{\xi}_E$. However, most importantly one must remember that this is only a technical assumption, allowing easier mathematical manipulations. Also in our presentation we assume the simplest ``passive'' description of the environment devoid of any internal dynamics. An example of decoherence model with ``active'' environment with self-evolution can be found in the article \cite{Horodecki22}, which is one of the contributions of the Jubilarian to the decoherence theory.

To the system and pointer of the von Neumann approach, which as we shall see now is rightly termed ``pre-measurement", we add the inevitable interaction with an environment in an initial state $\ket{\xi}_E$.
The pre-measurement couples (correlates, entangles), via a suitably precisely chosen interaction with the pointer $P$ variable (observable), the eigenstates of the observable of the system $s$, which is to be measured,  The interaction controlled by the devices constructed/used by experimenters who are understanding the physics of the situation, is specified by:

\begin{eqnarray}
     &\ket{\psi}_s\ket{neutral}_P \ket{\xi}_E    \xrightarrow{\textrm{pre-measure}}
     \sum_lc_l\ket{b_l}_s\ket{r_l}_P \ket{\xi}_E.& \nonumber \\ 
\end{eqnarray}
The pointer should be constructed such that it does not interact with the environment  when in the neutral position. E. g., in the Stern-Gerlach device the particle's position serves as the pointer, see Fig. \ref{fig:sg}. Before the particle reaches the region in space where are positioned the detectors it does not interact with the detectors. The detectors are external to the quantum system. As a whole, with all wiring, etc., they form the environment $E$.

Next stage is the interaction pointer-environment. As it is stressed by Zurek. e.g \cite{ZUREK2022}, if the measurement process is to work properly the interaction pointer-environment should not affect the pointer eigenstates $\ket{r_l}_P$, but must leave an imprint on the environment. Thus for properly functioning devices the next stage goes as follows:
\begin{eqnarray}
\xrightarrow{\textrm{coupling-with-E+internal-process-in-E}}
     \sum_lc_l\ket{b_l}_s \ket{r_l}_P\ket{\xi_l}_E.& \nonumber\\
\end{eqnarray}
As we still cannot see with our eyes the location of system $S$, in order to have a good measurement resolution, we demand that the states of the environment $\ket{\xi_l}_E$ correlated with the end location of the particle $\ket{r_l}_P$, are \textit{almost} orthogonal:
\begin{equation}
    |\braket{\xi_k}{\xi_l}_E| \approx 0,
\end{equation}
for all $k\neq l$.
How close it is to zero? Let us assume that the detectors and their principal wiring consist of at least of $10^{20}$ atoms, and the interaction with the particle causes say a small change of the quantum state describing say $10^6$ of specific atoms in the ionization chamber via the avalanche process (an internal process in the environment), which results  in a very small change of the state vectors, say $|\braket{f}{e}_n|=0.99$, where $n$ numbers the atoms in question. $\ket{e}$ is their initial state vector, while $\ket{f}$ is the final one. We have  $\prod_{n=1}^{10^6} |\braket{f}{e}_n| \approx 0$, and the Wolfram Mathematica response is  ``General: $0.99^{1000000}$ is too small to represent as a normalized machine number; precision may be lost, Out.$=\{0.\}$". Thus the final states of the detector as whole are perfectly distinguishable. Note that we have made very many obvious  approximations, all of which work toward increasing the estimate of $|\braket{\xi_k}{\xi_l}_E|$. In real conditions they are distinguishable by a spark, a click, here but not there, a warming up of the detector, and in the end appearance of a specific number on computer's display. Note that additionally if just one state describing {\em one} atom of the environment additionally ``flips", that is for some $n=n_a$ we have $|\braket{f}{e}_{n_a}|=0$, the mathematical zero, then we have precisely $|\braket{\xi_k}{\xi_l}_E| = 0$. Note that in the case of ionization we indeed have $|\braket{f}{e}_{n_a}|=0$.

The detectors are very complicated expensive devices, consisting of many atoms, wiring, etc.; they are the ``sensor organs" of the environment. Their structure can be split into the \textit{active zone}, which interacts with the system$+$pointer, macroscopic in size and with controllable initial microstate, and  the \textit{deep environment}, microstate which is describing it uncontrollable (mostly with exception of the temperature), because of the zillions of atoms of which it consists. The deep environment responds to the processes in the active zone, and as a matter of fact broadcasts information about the processes, actively because it comprises the wiring, and passively as an imprint is left in it (e.g. heat excess). Because of these features, further down we split $\ket{\xi_l}_E$ into $\ket{r_l}_D \ket{\xi_l}_{Env}$, where $D$ stands for the active zone (of the detector), and $Env$ for the deep environment. The response of the active zone is mainly dependent on the state describing  the pointer, that is why it is indexed by $r_l$. 

 Thus the final stage of the unitary evolution describing the measurement process is specified by:
 \begin{eqnarray} \label{U-FINAL}
 &\ket{initial} _{s\otimes P \otimes D \otimes Env} =  \ket{\psi}_s\ket{neutral}_P \ket{neutral}_D\ket{\xi}_{Env}& \nonumber\\     &\xrightarrow{\textrm{measurement-process}}
     \sum_lc_l\ket{b_l}_s\ket{r_l}_P \ket{r_l}_D\ket{\xi_l}_{Env}& \nonumber\\ 
     &=\ket{final}_{s\otimes P \otimes D \otimes Env}.&   \nonumber \\
\end{eqnarray}
That is it. Nothing more.

Despite the unitary character of the process we reach the predictive goal of quantum mechanics: prediction of 
a certain probability distribution of the measurement results. The reduced density matrix, after tracing out the deep environment, reads 
\begin{eqnarray}\label{FINAL}
 &\trace_{Env} \ketbra{final}{final} _{s\otimes P \otimes D \otimes Env}=\varrho^{(final)}_{s\otimes P \otimes D}&   \nonumber \\
\label{FINAL}
 &=    \sum_l|c_l|^2\ketbra{b_l}_s\ketbra{r_l}_P \ketbra{r_l}_D.&  
\end{eqnarray}
That is we have a classical probabilistic mixture of all possible final states, $\ket{b_l}_s$, correlated with position states of the pointer variable $\ketbra{r_l}_P$ and activation of detector signalling result $r_l$,  given by $\ketbra{r_l}_D$. As quantum mechanics gives only probabilities of specific results we cannot expect anything more. All that describes the operationally accessible events in the lab.  The description is for  a statistical ensemble, that is a conceptual representation of an infinite number of 
equivalently performed ideal experiments, or infinite number of totally equivalent identical independent experiments done at the same time. Testing this prediction in the lab one must resort to a frequentist interpretation of probabilities, or use statistical methods, which can be argued to be equivalent to that. For QBists $|c_l|^2$ are optimal betting strategies for the results in the lab. In the lab, in turn,  many, but still a finite number, of repetitions of the experiment are done with the hope of no systematic errors, e.g. due to the drifting in time of some parameters of the apparatuses. Note that in the above discussion we have just presented the basic logical structure of the interactions present in decoherence theory. The readers interested in a more detailed discussion of requirements for the system-environment interaction to lead to proper quantum measurement and information transfer about its result can familiarize themselves with \cite{KORBICZ-2014, Horodecki15} or \cite{Schlosshauer.04} and the most recent and comprehensive \cite{ZUREK-book-2025}.

Thus above sketched quantum measurement theory fully reflects the testable aspect of quantum mechanics as a set of physical laws.
As the description is for a statistical ensemble, there is no wonder that for each  element of the ensemble one expects a specific $r_l$ to pop up.  The ``third measurement problem", namely the question what decides about a \textit{particular} outcome,  {\em within} quantum theory (of measurement) does not exist, as the only quantum mechanical predictions are probabilities. Individual results of single runs/trials are an inherent feature of {\em any} probabilistic theory, which is a candidate to describe  measurable natural phenomena. 

\section{Wigner's Friend of Deutsch, and other daemons}
\label{sec:Wigner}
The ``unitary" quantum mechanics is essentially a modification of quantum mechanics which ends its discussion of the measurement process at the level of pre-measurement, but still allows specific outcomes $r_l$ to occur. These outcomes are produced with probabilities specified by the Born rule.

In contrast with the original quantum mechanics, the one before the quantum theory of measurement, the irreversible state update after obtaining a measurement result is not discussed.  No  theory of effectively classical behaviour of measuring apparatuses is introduced.
\begin{itemize}
    \item 
    One could single out the defining aspect of the ``unitarian" approach: it is the unacceptance of the irreversibility of the measurement. 
\end{itemize}

This, when translated to quantum mechanics  --  which describes classical behavior of macroscopic objects as a result of the decoherence process -- and thus allows to formulate the quantum measurement theory, {\em as we shall argue},  in effect transforms to insistence of the followers of unitary quantum mechanics that any unitary interaction can be in principle reversed.  Thus, also the measurement interaction presented above can be reversed\footnote{More generally, in unitary quantum mechanics there is no limit on human control of unitary transformations. See also Appendix A1.}, {\em and thus measurements are reversible}. Yes, mathematically any unitary transformation has its inverse, but in a laboratory one cannot reverse the unitary transformation leading to measurements, see the previous section.  

Note that this possibility of reversing the measurement, or equivalently (see appendix) to measure in a basis containing states describing the entire lab like $\ket{final}_{s\otimes P \otimes D \otimes Env}$,    is the very basis of the Deutsch's version of Wigner's Friend gedanken experiment \cite{Deutsch.85}. This work is a kind of a defining one for unitary quantum mechanics.
The controversial problem of Wigner's Friend has been  revived in  a recent discussion of quantum theory by Brukner \cite{brukner2015, Brukner17, Brukner.18}. A claim that ``quantum theory cannot consistently describe the use of itself" \cite{Frauchiger.18} is based on the idea that one can make measurements on entire labs. Also, despite a  verbally declared positive attitude to   decoherence theory, the Relational Quantum Mechanics (RQM) \cite{Rovelli.96} is essentially based on this, see its recent defense which overtly uses inversion of measurement interaction \cite{Cavalcanti_2023}.

\begin{figure}
    \centering
\includegraphics[width=0.9\linewidth]{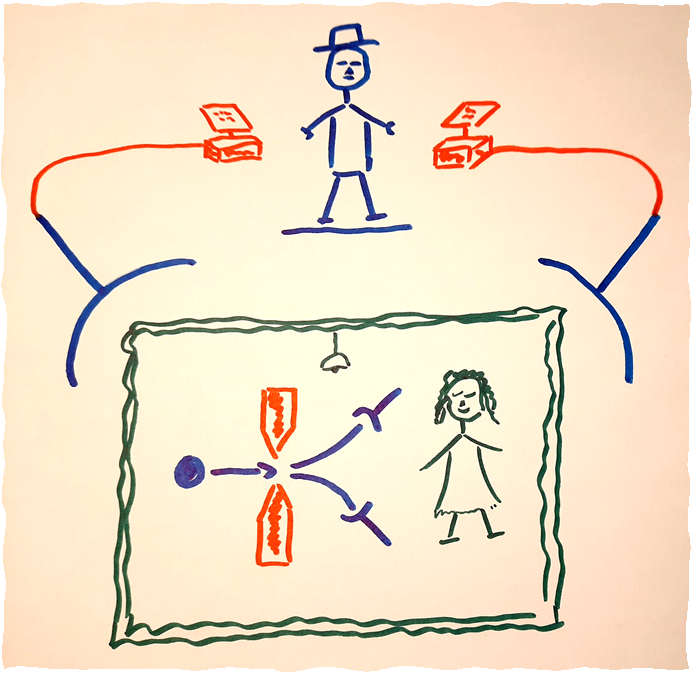}
    \caption{An illustration of a thought experiment called an \textit{extended Wigner's Friend scenario}. One observer, a Friend, is sealed in a lab and performs an experiment on some quantum system, e.g. a Stern-Gerlach experiment on a spin-half particle. An external observer, customarily called Wigner, has an ability to perform arbitrary measurements on the entire Friend's lab, including ones which project onto macroscopic superpositions of lab states, see eq. \eqref{eq:wf}.}
    \label{fig:wf}
\end{figure}

The mentioned works, as well as their numerous followers, e.g. \cite{Bong2020, Barrett22, Leegwater2022, Wiseman23, Haddara23}, discuss what is now called \textit{extended Wigner's Friend scenarios}, see an illustration in Fig. \ref{fig:wf}. For a recent review  see \cite{Schmid24}. The defining feature of such scenarios is the ability of an external observer, Wigner, to perform an arbitrary unitary transformation  on the entire lab of his Friend including her, which enables him to perform a measurement on the entire lab \textit{and her} in an arbitrary basis, or to reverse her measurement.

{\subsubsection{{\em Reversal of the measurement interaction} after the measurement is completed {\em is a fundamentally untestable hypothesis} and as such cannot be considered as a candidate for a physical law.}

The paradoxes obtained within ``unitary" quantum mechanics, e.g. \cite{Frauchiger.18},  paradoxically additionally support the claim in the headline of this subsection, as they reveal internal inconsistencies which haunt  this hypothesis. Also, claims, e.g. in \cite{Frauchiger.18}, that it is in principle possible to make a measurement in e.g. basis containing states like:
\begin{equation}
\label{eq:wf}
    \ket{\pm}_{s\otimes F}= \frac{1}{\sqrt{2}}\big(\ket{+1}_s\ket{+1}_F \pm  \ket{-1}_s\ket{-1}_F\big),
\end{equation}
where $F$ is essentially $P\otimes D \otimes Env$, that is Wigner's Friend and {\em her lab}, fall into this category of statements (see Appendix A). This is because to go from one orthonormal basis to another one performs a unitary transformation, and in the above case the feasibility  of this operation hinges on a hypothesis that one can have a precise operational control of the entire system $F=P \otimes D\otimes Env$. Only such a precise control would allow a reversal of the measurement interaction. Additionally, the supporters of ``unitary'' quantum mechanics are mute about what happens with the measurement results when measurement interaction is reversed.

\subsubsection*{Untestability of reversal of measurement interaction}
First we shall give an argumentation for Friend and her apparatus. 

One can estimate from below the number of atoms in her nervous system assuming that its mass is 1 kg, this can be divided by the mass of one mole of carbon atoms, which is 12 g. We get 80 moles, so we can approximately say the number of atoms in her nervous system is of the order of $10^{26}$. 
A joint operational control of such a number of atoms will never be possible.

A simple argument is as follows. Recall the ultimate dream of quantum technologies: the universal  quantum computer. Let us consider  non-universal device which would be capable of performing just one thing: a unitary transformation on $10^{26}$ qubits and its inverse. To be able to exactly perform such an operation the quantum computer/simulator must have in it installed at least $10^{26}$ qubits, and we forget here about error correction, etc. Such a quantum simulator could be used to test the possibility of performing and reversing a unitary transformation of a system which is described by a Hilbert space of dimension $\dim {\cal H} = 2^{10^{26}}$. Note that currently we are not able to perfectly inverse a polarization transformation on a single photon (argument: all experiments done so far have a final interference visibility $V<1$). Further, we know very well that a Mach-Zehnder interferometer is capable to perform any $SU(2)$ transformation. To this end, one must have a maximally precise control of {\em three} macroscopic parameters controlling the interferometer: essentially, relative phase shift at the input beams, relative phase shift in the internal beams, and relative phase shift in the exit beams, see Fig. \ref{MZ}. Note that this is related with three Euler rotations of a Cartesian basis of the Bloch Sphere. How many macroscopic parameters must one precisely control in a (science-fiction) interferometer performing an $SU\left(2^{10^{26}}\right)$ transformation? Note that the real dimension of the group $SU(n)$ equals $n^2-1$.

\begin{figure}
    \centering
\includegraphics[width=\linewidth]{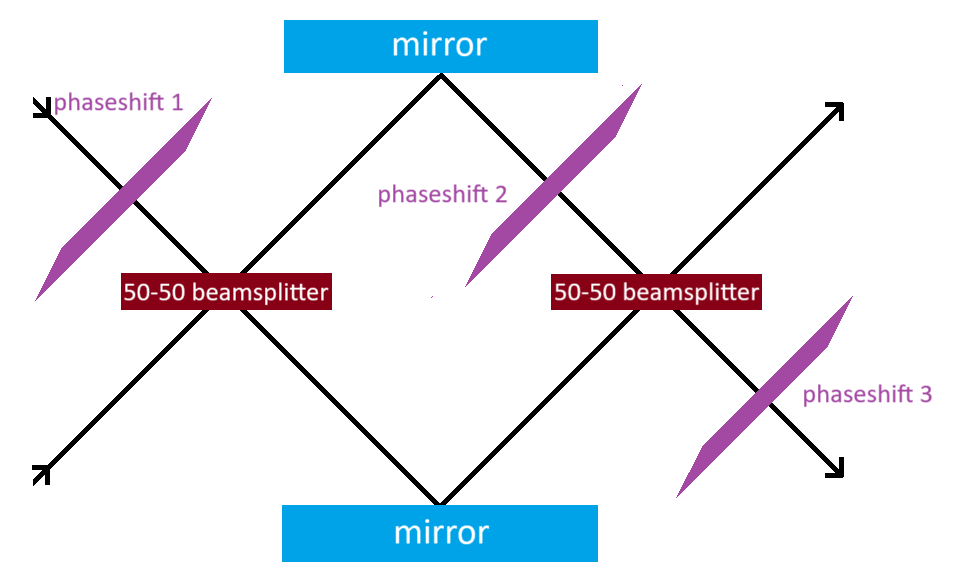}
    \caption{Schematic representation of a Mach-Zehnder interferometer, comprising two symmetric beamsplitters, two mirrors and three phase shifters, which serve as three parameters of the $SU(2)$ group. Note that typically one removes the first phase, as for single-photon input state it can be treated as a global phase.}
    \label{MZ}
\end{figure}

Assume now that only one neuron is to be simulated. Its mass is around $10^{-6}$ g, which gives $10^{-7}$ of a mole, and thus the number of atoms in it is, say, $10^{16}$.  Let  us wrongly assume that only one billionth of these atoms are relevant, and that they basically do not interact with the rest of the neuron: this gives us $10^7$. Thus a ``small" quantum simulator, in which complicated things like atoms, each described by an infinity dimensional Hilbert space,  are each replaced by a qubit, would have to be able to precisely perform and undo a unitary transformation in Hilbert space of dimension $2^{10^7}$. A similar argumentation could be given concerning simulation of a reversal of a measurement interaction within a detector. All these numbers are definitely beyond the Heisenberg Cut.\footnote{The quest for building a quantum computer is effectively a research attempting to push the Heisenberg Cut toward more and more complicated systems.} The tests which were discussed above are clearly impossible. They are even beyond science-fiction\footnote{Note that the number of particles in a visible universe is estimated to be \textit{only} of the order of $10^{80}$. As summarized by Streater in his book \cite{Streater2007}, chapter 4.6, \textit{``The problem is not with quantum mechanics, but cosmology''.} Therefore one can define statements in physics, which in order to be tested would require quantum computational powers beyond any capabilities of a planetary civilization, as belonging to metaphysics. This constitutes the ultimate Heisenberg's Cut. }.

All these estimates forget about interaction with the electromagnetic field, even with its vacuum, which can never ever be switched off. Forget, that the carbon atom has 12 electrons, and that its full description requires an infinitely dimensional Hilbert space. Forget, that during measurement interaction in say avalanche photo-diode, one has not only a generation of electronically detectable current, but also emission of photons, which simply fly away, and are gone forever. 

\subsubsection*{Can Friend inverse her own measurement?}

1. If this is done by inversion of the Hamiltonian, and allowing it to act via exactly the same time as its action during the unitary interaction transformation leading to measurement unitary transformation, $\hat{U}$,  then we face the following problem.
Friend sees a specific result, say $r_l$ thus for her the starting point of her reversal of the interaction is, 
\begin{equation}
\ket{b_l}_s\ket{r_l}_P\ket{r_l}_D\ket{\xi_l}_{Env}.
\end{equation}
Thus, applying $\hat{U}^{-1}$ would definitely {\em not} lead to the state $\ket{initial}_{s\otimes P\otimes D \otimes Env}$, compare (\ref{U-FINAL}).

2. Well, Friend may try to apply a unitary transformation $\hat{U}_l$ which has the following, definitely required, property, namely 
\begin{eqnarray} \label{U-L REVERSAL}
&\hat{U}_l\ket{b_l}_s\ket{r_l}_P\ket{r_l}_D\ket{\xi_l}_{Env}
&\nonumber \\
&=
\ket{initial}_{s\otimes P\otimes D \otimes Env}& \nonumber \\
& =  \ket{\psi}_s\ket{neutral}_P \ket{neutral}_D\ket{\xi}_{Env}&
\end{eqnarray}
But this cannot be done by a miracle, she must have a unitary transformation device (e.g. a super complicated interferometer)
which performs this evolution. This is a specific hardware with zillions of macroscopic settings controllable by her. The hardware and settings  are specific  for $\hat{U}_l$. For $l'\neq l$ one needs at least different settings, if not a different device. An agent in quantum mechanics must have a control over the settings. After the undoing of the transformation, Friend does not remember the results, see (\ref{U-L REVERSAL}), but she can look at the settings of  her super-measurement-inverter.  If the settings  are such that $\hat U_l$  was executed, then she finds out that the result was $r_l$..., thus the irreversibility still holds.

\subsubsection{Cathy and Erwin lurking in {\em Quantum Theory Needs No ‘Interpretation’} }

We shall comment here on a part of the important article \cite{fuchs2000quantum}: {\em Quantum Theory Needs No ‘Interpretation’}, which we quoted earlier, as it seems to be a manifesto of the approach we follow here.  It contains also a discussion the Wigner's Friend scenario, however, with a different wording.  This is the only part of the article that we deem as interpretative, against the wish of the authors. It is clearly based on unitary quantum mechanics, which we argue is an interpretation of quantum mechanics.

 The pure entanglement of system-Friend-(her lab) is treated as something obvious in the  manifesto of Fuchs and Peres \cite{fuchs2000quantum}.  We cannot agree with the following:
\begin{itemize}
    \item 
    \textit{``The
observer is Cathy (an experimental physicist) who enters her laboratory and sends a
photon through a beam splitter. If one of her detectors is activated, it opens a box
containing a piece of cake; the other detector opens a box with a piece of fruit. Cathy’s
friend Erwin (a theorist) stays outside the laboratory and computes Cathy’s wavefunction.
According to him, she is in a 50/50 superposition of states with some cake or some fruit
in her stomach. There is nothing wrong with that; this only represents his knowledge
of Cathy.''} 

Note that the above is a different wording of the Wigner's Friend paradox.
\end{itemize}

Sorry, but this does not represent Erwin's full knowledge of Cathy, photon, detectors and cake/fruit. We shall use here {\em reductio ad absurdum}. To {\em wrongly} ascribe {\em this} specific superposition (entanglement) Erwin must know Cathy's measurement basis. If he does not know the basis, but knows that she is going to perform for herself a measurement, it is for him a measurement in a random (but let us assume fixed) basis, and he is not able to (wrongly) ascribe the mentioned specific pure superposition (entangled state). 
Thus Erwin must know the measurement basis of Cathy, and that she for sure performed the measurement (according to the prior agreement). 

There is no reason for him to disregard this knowledge in his description of the quantum state inside Cathy's lab. If she did not measure or he is not sure that she did, there is no basis whatsoever to (wrongly) ascribe the specific entangled state implied by the text. {In such a case, Erwin has no other tools except for the full state tomography (on a sufficiently big ensemble of equivalent situations).} 

Therefore let us assume that Cathy is honest, and indeed performs the measurements, and she receives one of the possible  results (outcomes). Then the only thing that he does not know is Cathy's specific result. 
Thus, if unitary quantum mechanics formalism is to be applied to describe the situation in her sealed laboratory, this will be a grand entangled state describing her and the entire rest of her laboratory. 
\begin{itemize}
    \item However this is without any consequences, as this involves entanglement with the uncontrollable deep environment in her lab, neither controllable by her nor by him. It is impossible for him to control the micro-states of the environment. 
\end{itemize}

Thus, the only operationally relevant description is a mixed state with  probabilities (which are its eigenvalues) as defined by his quantum-formalism-based assessment of the probabilities of Cathy to get specific results (these are the ones calculable with Born rule). This would describe the classical correlation of the photons, the detectors, and her records. Paraphrasing the quotation: {\em Cathy’s
friend Erwin (a theorist) stays outside the laboratory and computes the quantum state describing Cathy’s situation.
According to him, she is in a 50/50 classical probabilistic mixture of states with some cake or some fruit
in her stomach. }

Note that such (wrong) picture (system-Cathy plus extra, entangled) is 
often lifted to the very essence of Qbism, see especially e.g. \cite{Mermin2016}, which is discussed in the Appendix \ref{app:mermin}. While we are very impressed by the Qbist analysis, because of the above reasoning, we cannot accept the supposed optimality of Erwin's betting in the discussed case.

If one {\em additionally} accepts that Erwin has powers to reverse the unitary transformation that happened in Cathy's lab\footnote{Cathy cannot do it for him by inversing the unitary transformation of the interactions, because she witnessed a specific result, therefore she ascribes a specific post-measurement state to herself-photon-detectors-cake and not the superposition. To put it short, for her the system is not entangled with her.}, which will never be possible as we showed earlier, he must open her lab, which is a brutal interaction (imagine the state of a tin can before and after opening it...). This leads to an immediate interaction of Cathy's environment with Erwin's environment, then as we argued, he is also not able to perform a measurement in a basis which contains the  entangled state, as one of its elements. Note that the state would describe the entanglement of Cathy, the photon, the detectors, cake, fruit, {\em and the deep environment in all these
elements of her lab and her}.}

The (wrong) state assignment described above (the superposition) would lead to an (operationally impossible)  boring experiment in which he would get for the full ensemble of repetitions just one result, which is equivalent to a positive test of his  wrong  state assignment (wrong, as he was ignoring a certain part of his knowledge). Of course, for the correct state assignment (the mixed state) quantum mechanics predicts random results for any measurement involving  bases of this kind.

\subsubsection*{Summary of the argument}

A reader may check that our description of the situation is concurrent with the paper of Peres \cite{peres1986quantum} in which the Author stresses the following: \\

{\em ``A measuring apparatus must have
macroscopically distinguishable states, and the word “macroscopic” has just been
defined as “incapable of being isolated from the environment.” If there is no
irreversibility, there are no measurements.''}\\

\noindent We wonder why in \cite{fuchs2000quantum} one can find: {\em  "If
Erwin has performed no observation,
then there is no reason he cannot
reverse Cathy’s digestion and memories. Of course, for that he would need
complete control of all the microscopic degrees of freedom of Cathy and her
laboratory, but that is a practical
problem, not a fundamental one."}

{\em To summarize}. Friend/Cathy has done a measurement which is preparation of new states describing the system $\ket{b_l}$. In each run one of such states is prepared with statistical frequencies given by the Born rule. It is an irreversible proces, as both Friend and Wigner/Erwin do not have any control on the uncontrollable states of the deep environment of Friend's lab. Thus, when assessing the state describing the situation in Friend's lab Wigner is limited to the degrees of freedom (or subsystems) which are controllable. States describing  these give the predictions for his measurements, or for whatever action he takes. Note that when describing predictions of observable lab effects, for correctly working apparatuses, one does need, and is not able to specify all the miscrostates of the atoms forming their deep environment. There is simple analogy with Bell-type entanglement experiments. Bob in his lab does not have any control over the events and action of Alice who is in the other remote lab. Thus for Bob the states describing the (ensemble of) particles arriving in his lab is a reduced density matrix for his subsystmes. This in the case of Bell experiments, as they involve entanglement, is two or more element proper probabilistic convex combination giving a mixed state, and never a pure one. 


\section{Conclusions}
The aim of our article is to show why the irreversibility associated with quantum measurements is \textit{not} a possible to overcome practical problem, but a fundamental one because of the very nature of macroscopic measuring devices. Note, that in his classic book \cite{peres2002quantum} Peres spells out the basics of our approach (page 376):\\

``{\em Consistency
[...] requires the measuring process to be irreversible. There are no superobservers
in our physical world.}" \\

Let us summarize our considerations on measurement (ir-)reversibility:
\begin{itemize}
    \item 
    Any theoretical construction based on quantum mechanics which extends or completes it, but does not challenge its formalism and Born rule for real laboratory experiments/observations  giving classical signals/information on the obtained measurement outputs, is an \textit{interpretation} of quantum mechanics.  
That is, interpretations of quantum mechanics by definition cannot modify its predictions. Just as an interpretation of a literary text cannot modify the text itself. Interpretations add some meanings, ontic notions, whatever. As the interpretational  additions do not change the predictions concerning any laboratory experiment, they are fundamentally untestable, and thus they are metaphysics. The assumption of in principle reversibility of measurements is, as we show, operationally untestable.
    \item
    ``Unitary quantum mechanics'' differs from (standard, operational, orthodox, ``textbook", etc.) quantum mechanics in allowing the possibility of reversing measurements and in allowing the application of Born rule in
situations other than these which lead to macroscopically observable events. The possibility of reversing acts of measurement is in fact equivalent to the possibility of performing a measurement on entire quantum laboratories, including  experimenters.  Such predictions, which go beyond quantum mechanics, are not testable, and actions  to implement them are operationally impossible  -- this we show using the quantum measurement theory based on the quantum theory of classicality which uses the mechanism of decoherence.
Therefore we claim that {\em unitary quantum mechanics is an interpretation of quantum mechanics}, as it does not change  operationally accessible predictions of quantum mechanics.
As such it is metaphysics.\footnote{Note that, in contrast, local realistic models, which were thought to be a possible interpretation (completion) of quantum mechanics, thanks to Bell's theorem turned out to be a \textit{testable} modification of quantum mechanics, and thanks to the loophole-free tests of Bell inequalities \cite{Hensen2015}, \cite{Shalm2015}, \cite{Zeilinger2015} and \cite{HARALD}, were falsified as a possible description (law) of Nature.}
\item
  The clinical examples are  various versions of the Deutsch Wigner's Friend gedanken-experiment \cite{Schmid24} analyzed with the use of unitary quantum mechanics. In various versions of such an analysis of the Wigner-Friend situations the operationally impossible  reversal of measurements is used, or the possibility of performing a measurement on entire quantum laboratories is postulated.  
\end{itemize}

In quantum mechanics itself  there is no ``measurement/collapse problem". The only prediction that quantum mechanics gives are the probabilities (via the Born rule). These probabilities are laboratory tested/testable by relative frequencies of measurement events in statistically relevant number of repetitions of the measurements. Thus, the wave function/quantum state, from the point of view of probability theory describes the statistical ensemble of equivalently prepared quantum systems. Via the Born rule quantum state allows to predict all possible probabilities of all possible laboratory events. Note that measurement results happen in single runs of a quantum experiment. Quantum formalism applies to and describes only  ensembles of such runs. It is not a theory of single runs. A specific observed event in the final ``measurement" stage of a run of an experiment, if the measurement is non-demolishing, is a preparation of a member system  of newly defined  statistical ensemble of systems, which gave this and not other result, and as such is described by a new ``quantum state". 

The last remark: note the following apparently paradoxical situation. All extended Wigner-Friend paradoxes can be on a conceptual level squeezed to the following statement (see \cite{Schmid24}, Sec. III): \textit{the following three assumptions cannot be jointly true: (i) absoluteness of observed events, (ii) possibility of applying and reversing arbitrary unitary operation on any system, and (iii) validity of Born rule for relative frequencies of measurement outcomes.} On a formal level we agree that these three properties cannot be jointly true. However, the difference between us and the followers of unitary quantum mechanics is  in the way one treats the Born rule, and in our insistence that measurements are irreversible. According to adherents of unitary quantum mechanics Born rule can be meaningfully applied to ``outcomes'' somehow assigned to a unitary processes, whereas we claim, that it can be meaningfully applied solely to relative frequencies of macroscopically observable events of the kind used to define preparations of the quantum states, that is measurement results. Measurements are fundamentally irreversible, because of the macroscopic nature of measuring apparatuses and the fact that they are to produce classical, that is cloneable, information about the results. This is despite the fact that one can formally describe the interaction between system and apparatus as a unitary transformation. Therefore, one can say that the extended Wigner-Friend arguments, which use ``unitary quantum mechanics" in any version, can be invalidated just by looking at the origins of quantum mechanics. 

\subsection{Decoherence-based quantum measurement theory reproduces operational quantum mechanics}
All that we wrote boils down to a recipe: ``shut up and calculate" the  observable effects. This approach is used by many a scientist who are not bothered with the interpretational issues. But only those who do not treat various visualizations seriously, even if the visualizations help them in their work. Often, if the logical consequences of such visualizations  are precisely analyzed they can usually be shown to lead to a kind of internally inconsistent attempts to build an interpretation of quantum mechanics. We stress the word \textit{attempts}, because we do not deny existence of internally consistent interpretations such as Copenhagen, Bohmian, Many Worlds, etc.,
and even of the ``unitary quantum mechanics".

We do not introduce any new approach to the decoherence theory of quantum measurement. We want to put forward that 
one may look at the decoherence theory as it is, as the road toward a finalization, if one hesitates to use the word completion, of Bohr's program (see e.g. \cite{bohr1996discussion}). His intuitions get a firm foundation (for a concurrent view see  \cite{CAMILLERI201573}). The pointer variables of macroscopic classical apparatuses behave  effectively classically because they comprise zillions of atoms, which allows observable values/outcomes/results and pointer positions to be, in theory, in perfect correlations, consistent with Born rule.
The fundamental randomness, and complementarity  remain as the features of the theory. We used here the phrase ``Bohr's program", as it seems to us that he wanted to point toward a specific way of thinking about quantum mechanics, rather gave directions, and did not attempt to set unquestionable axioms beyond which one cannot go. Maybe he wanted that, but he never achieved. The various interpretations that emerged seem do confirm this opinion.

Many interpretations of quantum mechanics aim at understanding the roots of the quantum randomness. Essentially they treat it as a weakness of the theory, see e.g. the flagship paper of this tendency by Einstein, Podolsky and Rosen \cite{EPR.35}. The state is supposed to be an ontic or epistemic feature of a quantum system often supplemented with hidden variables explaining the emergent randomness. Zurek advocates for "epiontic" status of the states, see \cite{ZUREK-book-2025}.

We suggest another road, which seems as operational as it gets, and is fully consistent with quantum formalism and with the quantum measurement theory based on the decoherence theory of classicality:

\begin{itemize}
    \item 
    The statistical ensemble of equivalently prepared quantum systems, if there is no element of classical randomness, is described by a vector in a Hilbert space of dimension equal to the maximal number of fully distinguishable measurement results which are possible for the quantum systems. Note: no ontic or epistemic direct relation with the systems, just a comprehensive representation of the initial macroscopic preparation defining the statistical ensemble.
    \item 
    The evolution of the description of the preparation of the statistical ensemble is unitary, and deterministic (motion from one vector in the Hilbert space to another, a continuous one: Schroedinger equation.)
    The unitary evolution applies to statistical ensembles of the systems to be measured and of the ``atomic" structure of the measuring apparatuses, as well as their interactions.
    \item 
    When one ignores (mathematically: traces out) all macroscopically irrelevant, uncontrollable and inaccessible, during the specific measurement, ``degrees of freedom" of the apparatus (which form an internal "deep" environment), Born rule emerges for the statistical ensemble as the classical probabilistic correlation of measured observable-values with the macroscopic pointer "position" (often a collective variable) of the apparatus.
    \item For pointers one chooses degrees of freedom of the apparatus which can couple with the  system not affecting the measured observable (technically: the unitary interaction of pointer with the observable commutes with the observable). The interaction pointer-internal environment should also  effectively not disturb the pointer observable. 
\end{itemize}
This set of rules is in agreement with the operational quantum mechanics, and with Bohr's insistence that macroscopic measuring apparatuses must be (effectively) classically describable. Note that the classical describability of measuring devices is limited to some macroscopic parameters (variables, degrees of freedom), and all microstates of the atomic micro structure are ignored -- exactly in the same way as we have in the case of the statistical theory based on the hypothesis of atoms applied to thermodynamics. Note further, that before the decoherence theory many interpretative approaches were based on supposed 
analogy between the macroscopic (collective) thermodynamic parameters (as ignoring micro-states of "atoms") and quantum states (as ignoring hidden variables, hidden causes, etc.). Within the decoherence theory we ignore microstates describing the atomic structure of the {\em measuring devices}, by tracing them out, while retaining the description of collective degrees of freedom which interact with measured systems, and have an effective classical description (after the decoherence process).

Paradoxically for the adherents of unitary quantum mechanics, we have only unitary evolution of the description of statistical ensembles of quantum systems and (in their internal structure: quantum) measuring apparatuses. \textit{There is no collapse understood as a physical process}.
Measurements split the statistical ensemble into subensembles defined by the specific result. The subensembles are equivalent to new preparations.

At the 2025 Helgoland conference Anton Zeilinger reportedly said “The quantum state only describes our knowledge.” This can be found in an article in Physics Magazine by Henderson \cite{Henderson25}. We think that our essay is concurrent with this view.

\section*{Acknowledgements}
MZ thanks Ad{\'a}n Cabello for insisting that an essay like that should be written.
A private communication by our collaborator Jay Lawrence, after his reading of the first sketch of this article,  that ``measurement is *logically* irreversible", might have influenced some of our thoughts, as the timing of the  correspondence does not exclude a causal link.
This work is supported by the IRAP/MAB programme, project no. FENG.02.01-IP.05-0006/23, financed by the MAB FENG program 2021-2027, Priority FENG.02, Measure FENG.02.01., with the support of the FNP (Foundation for Polish Science).

%

\appendix

\section{Covering some questions that the Reader may have about some technicalities}

Here we give our answers to some questions which we hear during discussions. The list might be expanded in future updates of the manuscript. Each subsection forms a separate unit, related with the main text, usually being an additional explanation, but rather unrelated with other subsections here.

\subsection{Control of unitary transformation, and measurement} \label{F}

Adherents of unitary (interpretation of) quantum mechanics, especially when discussing the Wigner's Friend paradox, employ in their reasoning either reversal of the (unitary) measurement interaction involving $P\otimes D \otimes Env$ (that is Wigner's Friends and her lab), or measurements containing basis states of the form
 \begin{equation}
    \ket{\pm}_{s\otimes F}= \frac{1}{\sqrt{2}}\big(\ket{+1}_s\ket{+1}_F \pm  \ket{-1}_s\ket{-1}_F\big).
\end{equation}
 We shall show below that these are two faces of the same coin.

The essence of co-existence of Schroedinger and Heisenberg pictures in quantum mechanics is that an evolution can be equivalently described as an evolution of the state or evolution of the observables. Both pictures involve the same unitary operator. This is analogous to the passive and active view of rotations of Cartesian coordinates' basis vectors in Euclidean space. In mathematical terms  measurement on an ensemble described by $\ket{\psi}$ in basis $U\ket{B_i}$ is equivalent to measurement in basis $\ket{B_i}$
and the system evolving back to $U^{-1}\ket{\psi}.$ This shows that  a control of the full reversal of a unitary interaction leading to measurement (with all that interaction with the environment), allows one to perform measurement in a basis containing the state $\ket{final} _{s\otimes P \otimes D \otimes Env}$ of the process described in (\ref{U-FINAL}). 

\subsection{Remark on pure states and preparations}\label{ENSEMBLE}
If we assume that we do  not know what is meant by a pure state in quantum mechanics, but we know the mathematical structure of quantum mechanics including the Born rule, then, using just logic one can show that the state encodes a preparation of a specific statistical ensemble. 

Take state 
 $\ket{\psi}$, it is an eigenstate  of an observable $\Pi_\psi=\ketbra{\psi}{\psi}$. In the earlier ``orthodox" approach to measurement this means that the state can be prepared by measuring
 observable  $\Pi_\psi$, and selection of only the instances when the outcome value was $1$. Thus, this is one of the  methods to prepare the state (there are  many other methods, but this selection-by-measurement method is one of them). 
 
 Still, we might think that $\ket{\psi}$ could be associated with {\em each} of the systems, which in the measurement of $\Pi_\psi$ gave as a result value equal to eigenvalue  $1$. Manyfold repetition of this procedure obviously would constitute an ensemble of equivalently prepared particles.
 However, one can reverse the situation. Consider a sequence of particles that somebody prepared for us in state $\ket{\psi}$, but forgot to tell us that the state is $\ket{\psi}$. In such a case there is no way to find out ``in which state" is a single particle, if we are to measure just one of these. However, a tomographic experiment on the ensemble would give an approximate answer in real experiments, and in theory a gedanken version of it gives the exact answer as then one uses the abstract statistical ensemble.

 \begin{itemize}
     \item  Thus $\psi$ definitely gives the full quantum description of an ensemble of equivalently prepared systems. But the relation of  a single system with $\psi$, after the aforementioned ensemble preparation, is only that the system belongs to the ensemble. Statements like ``system is in the state $\psi$" have an interpretational character as they have no justification in operational procedures. They face the criticism of Einstein mentioned in \cite{Schilpp59}, see the itemized quotation in subsection \ref{PSI}. If one insists that this is literately the case then we enter metaphysics.
 \end{itemize}

 Note finally, that every feasible preparation method gives us an ensemble of systems described by a pure state or a mixed state. However, every $\ket{\psi}$ or $\rho$ describes the result of a theoretically possible preparation of an ensemble, but does not give us a recipe for a feasible preparation.

\section{Covering some questions that the Reader may have about relation of all that with specific interpretations}
\subsection{Copenhagen}

The Copenhagen ``interpretation" has many versions, some may even overlap with what we present as operational quantum mechanics. The versions of it in which one claims that {\em  ``the (individual) system is completely described by its state vector  $\ket{\psi}$"}, or something equivalent, contain for us a sufficient  reason to classify such a Copenhagenish  formulation of quantum mechanics  as an untestable interpretation. As quantum mechanics is a probabilistic theory, it is mute about states of individual systems; the ket $\ket{\psi}$ represents mathematically a preparation procedure of an ensemble (of systems, or runs of an experiment, see the main text), see subsection \ref{ENSEMBLE}. Note however, that such versions of the Copenhagen interpretations have one positive side: they do not introduce any new elements to the quantum description. We have no many worlds, no non-local hidden variables (Bohmian approach), no relative facts (Relational Quantum Mechanics), etc. - see further. 

Our approach can be classified as Copenhagenish if one declares that: \textit{a statistical ensemble of equivalently prepared systems is completely described by the ket vector  $\ket{\psi}$ which allows us to get, via the Born rule, all possible probabilistic predictions  for future measurements. No other additional description is available for individual systems that belong to such an ensemble.  } 

Note that, as we argued, non-pure density operators  provide a description that can be completed \textit{within} the quantum theory. That is such a completion does not require new descriptive elements or variables. Either we have some classical randomness in the preparation procedure, or we have a preparation that gives entangled states of at least pairs of systems, and we want to have a description of the ensemble of one of the systems from entangled pair.

\subsection{Ballentine}
Ballentine's review \cite{Bal70} contains statements that go beyond claimed therein meaning of the quantum states as a mathematical description
of the ensemble of equivalently prepared systems. Quotation from page 361 reads: ``\textit{For example, the system
may be a single electron. Then the ensemble will be the
conceptual (infinite) set of all single electrons which
have been subjected to some state preparation technique (to be specified for each state), generally by
interaction with a suitable apparatus. Thus a momentum eigenstate (plane wave in configuration space)
represents the ensemble whose members are single
electrons each having the same momentum, but distributed uniformly over all positions.}" In quantum mechanics such an ensemble gives a prediction that if one decides to measure position on systems which belong to this ensemble, the results would be absolutely random. However the state (i.e. ensemble) preparation procedure is mute about positions at the stage of preparation. Phrases like ``\textit{but distributed uniformly over all positions}'' of the quotation above are of an interpretative character. Quantum mechanics tells us only that the description of the ensemble of equivalently prepared systems -- the ``state'' -- does not define their positions.
Thus, the approach presented in our article differs from  the one of Ballentine. 

The remark of Ballentine is not an accidental imprecision, as one finds concurrent statements in the entire review. Still, as Ballentine calls his approach ``statistical interpretation'', this is consistent with our understanding of interpretations, since interpretations must contain some untestable element which goes beyond the quantum formalism.


\subsection{Relational Quantum Mechanics (RQM)}
    Relational Quantum Mechanics (RQM), originally proposed by Rovelli in 1996 \cite{Rovelli.96}, and further developed in several extensions, see e.g. \cite{Adlam.22, Biagio.22}, positions itself as an interpretation of quantum mechanics with a specific ontology (realistic and naturalistic according to its founders, see first paragraph of \cite{Adlam.22}) based on \textit{relative facts}. To understand this notion one has to know that RQM identifies an interaction between any two quantum systems as a measurement. Outcomes of such measurements (called by us \textit{RQM measurements}, in order to distinguish them from ordinary quantum measurements) are called relative facts, and are accessible only to the systems involved in the interaction \footnote{The \textit{relative} character of outcomes in RQM is motivated by the following supposed analogy with special relativity (\cite{Rovelli.96}, p. 1): ``\textit{As such, it bears a vague resemblance
with Einstein’s discussion of special relativity, which is
based on the critique of the notion of absolute simultaneity. The notion rejected here is the notion of absolute, or observer-independent, state of a system; equivalently, the notion of observer-independent values of physical quantities. The thesis of the present work is that by abandoning such a notion (in favor of the weaker notion
of state –and values of physical quantities– relative to
something), quantum mechanics makes much more sense}''. In our opinion such presentation is misleading. In special relativity one indeed has a relativity of simultaneity for spatially separated events, but objectiveness of events holds in an unwavering way: if something happens in some point in spacetime, it is an objective event. What is relative is the transfer of information about the values registered at some point in spacetime to another observer, hence different observers can associate different states to describe the same system. For a specific  example see the standard quantum teleportation protocol described in Sec. \ref{sec:teleport} and in the Fig. 2.}. 

In RQM no von-Neumann-L\"udders state update is used by external agents who are not involved in the unitary interaction which leads to  the RQM \textit{measurement}, and which would lead to a post RQM measurement description of the situation in terms of a mixed state which is a classical probabilistic convex combination of all possible states describing specific outcomes.
    This feature is readily apparent in the RQM description of the Wigner-Friend scenario, in which it is assumed that the Friend $F$ obtains a relative outcome due to a unitary interaction with the system $S$, and at the same time Wigner, who does not interact with Friend's perfectly isolated lab $L$ assigns an entangled state to the composite system $L\otimes F\otimes S$. In a recent work \cite{Lawrence22} we have shown that such understanding of a quantum measurement is at odds with quantum complementarity and leads to a contradiction. 
    
     In a recent paper \cite{Adlam.22} followers of RQM add a modification called \textit{cross-perspective links axiom}, which, in simple words, states that if Wigner measures Friend's degrees of freedom in which her relative outcome has been encoded, he would obtain outcome consistent with Friend's original relative outcome. This, according to RQM proponents, ''\textit{ensures that observers can reach inter-subjective agreement about
quantum events which have occurred in the past, thus shoring up the status of RQM as a form of scientific realism and ensuring that empirical confirmation is possible in RQM}``, see Introduction in \cite{Adlam.22}. Unfortunately, as we show in \cite{Markiewicz23}, enrichment of RQM with cross-perspective links axiom makes the situation even worse, namely one obtains an internal contradiction  within RQM postulates, a conflict between the old ones and the new postulate.

Additionally the adoption of the cross perspectives links axiom makes RQM a hidden variable theory.  The result obtained by Friend dictates the result of Wigner's measurement, if later the latter chooses to measure the system, or Friend's degrees of freedom in which her relative outcome has been encoded, in the basis of Friend's RQM measurement. This clearly is an additional description of the Friend-system, beyond the one given by the assumed in RQM Friend-system entangled state. {\em Friend's RQM outcome} is unknown to Wigner, but has a deterministic predictive power. Thus, this {\em is a perfect hidden variable}. It differs from usual hidden variables of other approaches by appearing only after Friend's RQM measurement, and out of the two actors only for Wigner it is hidden.

\subsection{ Qubism}
\label{app:mermin}
The current version of Qbism {\em now} includes Wigner's Friend in a superposition. This is explicitly stated in a recent Qbism's analysis of Wigner's Friend scenario (\cite{Qbism2020}, top of the page 9): \textit{It follows that a QBist can simultaneously assign the state $\ket{\Phi}$ and grant his friend a conscious
experience of having seen either “up” or “down”.}
$\ket{\Phi}$ represents there a superposition (entangled) state describing the particle-friend system.

We somehow feel that it was not the case in the {\em initial} exposition of the Qbist program, which can be found in  \cite{Caves02}. Our view is supported by the second paragraph of the Introduction in \cite{Qbism2020}, which emphasizes that Qbism has made a turn towards \textit{relativisation} of measurement outcomes\footnote{By this we mean, within the context of this discussion, that the results of Friend exist for her but not for Wigner. They are not merely unknown for him. This kind of relativisation we find internally inconsistent, but nevertheless we think we must present here this point of view of its adherents.}, with its intermediate step in the work \cite{Caves07} and final declaration, that \textit{In fact, Wigner’s friend was central to the development of QBist
thinking} \cite{Qbism2020}.

Finally, Mermin in his acclaiming discussion of Qbism overtly puts himself on the side of Deutsch in the case of Wigner's Friend problem \cite{Mermin2016}.

The current version of Qbism is therefore effectively a restricted form of Relational Quantum Mechanics of Rovelli, plus the Bayesian interpretation of classical probability.  Qbism accepts measurement results to appear due to entangling interactions only for complicated systems like Wigner's Friend. Relational Quantum Mechanics assumes that outcomes are results of any unitary interaction leading to an entangled state describing two systems. 

Therefore Qbism  has evolved to effectively a form of a hidden variable theory, because if Wigner performs a checking measurement then he ({\em deterministically}) receives Friend's result.  Note that the checking of the result can be done on the system only, as we have the quantum mechanical rule that a repeated measurement reveals the same result as the first one, and it does not matter who repeats it. The only requirement is that the system does not evolve between the two measurements. Additionally such a procedure would not change the state describing Friend's memory.

Thus Wigner's description of Friend's situation, in the form of her entanglement with the measured system is complemented by her (hidden for Wigner) result. {\em In quantum theory if a situation is described by a pure state, no additional description is allowed.}\footnote{Note that a proper mixed state allows additional description: either the given system described by it is in an entangled state with another system, perhaps unknown one, or the state is a classical probabilistic mixture of at least two pure states, which means that the source is not defining the state maximally, additional description is needed, like Alice playing with her  half wave plate.}

\subsection{Deutsch}
Deutsch gedanken experiment \cite{Deutsch.85} rests upon the reversibility assumption, thus the test he proposed will never be possible.
Thus, Many Worlds Interpretation, which the gedanken experiment was to support, will keep its (metaphysical) status of an interpretation. 

Note however the following surprising analogy between the ensemble approach and Many Worlds Interpretation, see e.g. \cite{peres1986quantum}, page 445.
As the theoretical concept of a quantum state, $\ket{\psi}$, applies to an infinite statistical ensemble of equivalently prepared systems, within the entire ensemble all possible results happen! Just as in the statistical ensemble of dice rolls $1, 2, 3, 4, 5$ and $6$ happen. In a perfect laboratory realization of an experiment,  a sufficiently large number of runs leads to a situation in which with a probability close to 1 all non-zero-probability results would happen.


\end{document}